# 5G Cellular: Key Enabling Technologies and Research Challenges

Ekram Hossain, *Fellow*, *IEEE*, and Monowar Hasan, *Student Member*, *IEEE*

Department of Electrical and Computer Engineering, University of Manitoba, Canada

Email: Ekram.Hossain@umanitoba.ca, monowar_hasan@umanitoba.ca

The evolving fifth generation (5G) cellular wireless networks are envisioned to provide higher data rates, enhanced end-user quality-of-experience (QoE), reduced end-to-end latency, and lower energy consumption. This article presents several emerging technologies, which will enable and define the 5G mobile communications standards. The major research problems, which these new technologies breed, as well as the measurement and test challenges for 5G systems are also highlighted.

**Introduction**

The exponential growth of wireless data services driven by mobile Internet and smart devices has triggered the investigation of the 5G cellular network. Around 2020, the new 5G mobile networks are expected to be deployed. 5G networks will have to support multimedia applications with a wide variety of requirements, including higher peak and user data rates, reduced latency, enhanced indoor coverage, improved energy efficiency and so on.

The primary technologies and approaches to address the requirements for 5G systems can be classified as follows [1-3]:

- densification of existing cellular networks with the massive addition of small cells and a provision for peer-to-peer (P2P) communication (e.g., device-to-device [D2D] and machine-to-machine [M2M] communication-enabled multi-tier heterogeneous networks);
- simultaneous transmission and reception (e.g., full-duplex [FD] communication);
- massive multiple-input multiple-output (massive-MIMO) and millimeter-wave (mm-wave) communications technologies;

- improved energy efficiency by energy-aware communication and energy harvesting;
- cloud-based radio access network (C-RAN); and
- virtualization of wireless resources.

The objective of this article is to provide a brief review of the emerging technologies that could shape the design of future 5G cellular networks. The key ideas for each of the technologies are stated, along with the potential impact on 5G requirements and the open research issues.

The article is organized as follows: followed by a brief overview of the networks and devices for 5G systems, a brief outline of the enabling technologies is presented. The key challenges related to measurement, testing, and validating the performance of 5G system components are briefly discussed. Then the fundamental research challenges for resource management in 5G systems are highlighted before we conclude the article. The visions and requirements of 5G networks and the corresponding technologies are presented in Table I.

TABLE I

FEATURES AND TRENDS OF 5G NETWORKS

| 5G expectations and features | Trends/proposals |
|---|---|
| Capacity and throughput improvement, high data rate (~1000x of throughput improvement over 4G, cell data rate ~10 Gb/s, signaling loads less than 1~100%) | Spectrum reuse and use of different band (e.g., mm-wave communication using 28~GHz and 38~GHz bands), multi-tier network, D2D communication, C-RAN, massive-MIMO |
| Reduced latency (2~5 milliseconds end-to-end latencies) | Full-duplex communication, C-RAN, D2D communication |
| Network densification (~1000x higher mobile data per unit area, 100~10000x higher number of connecting devices) | Heterogeneous and multi-tier network |
| Advanced services and applications (e.g., smart city, service-oriented communication) | C-RAN, network virtualization, M2M communication |
| Improved energy efficiency (~10x prolonged battery life) | Wireless charging, energy harvesting |
| Autonomous applications and network management, Internet of Things | M2M communication, self-organizing and cognitive network |

**Networks and Devices for 5G Systems**

The 5G networks will consist of nodes/cells with heterogeneous characteristics and capacities (e.g., macrocells, small cells [such as femtocells and picocells], D2D user equipments [UEs] etc.), which will result in a multi-tier architecture. Due to increasing complexity in network management and coordination among multiple network tiers, the network nodes will have the capability of *self-organization* (e.g., autonomous load balancing, interference minimization, spectrum allocation, power adaptation etc.) [2]. Also, a UE can have simultaneous active connections to more than one base station (BS) or access point (AP) using the same or different radio access technologies (RATs). The heterogeneous nodes (e.g., UEs, BSs, smart machines, wearable devices etc.) can be integrated through a unified (possibly cloud-based) network to provide seamless connectivity. The communication efficiency in 5G systems will be improved by incorporating techniques such as coordinated multipoint (CoMP) joint transmission and reception, network-assisted interference cancellation and suppression, spectrum reuse (e.g., non-orthogonal multiple access), and three-dimensional or full-dimensional MIMO [4]. In addition, the use of a large number of remote radio heads (RRHs) connected to central processing nodes (e.g., clouds) with the high-speed backhaul/fronthaul is also envisioned [3].

Due to the emerging trends of energy-aware FD communication and spectrum virtualization, 5G device architectures will be more complex than those with 4G systems. Devices in the 5G networks should be capable of operating in multiple spectrum bands, ranging from radio frequency (RF) to mm-wave, while being backward compatible (e.g., with existing technologies such as 3G and 4G). Due to energy hungry multimedia applications, energy efficiency will be an important feature for 5G user experience and hence it is desirable that the devices integrate energy harvesting technologies [3]. The need to support several RATs, FD communication and energy harvesting capabilities will breed many challenges for designing and testing the internal chips and front-end modules for 5G devices.

Fig. 1 illustrates the aspects of 5G networks and devices discussed above.

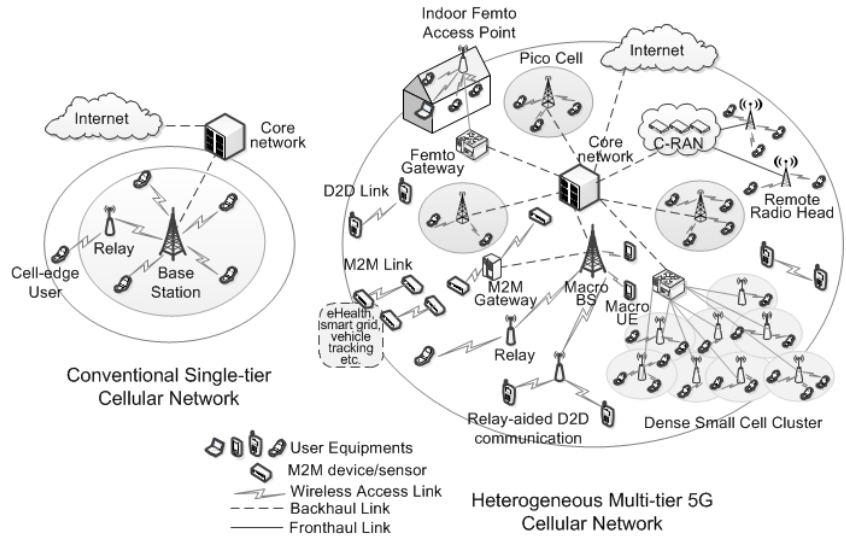

(a)

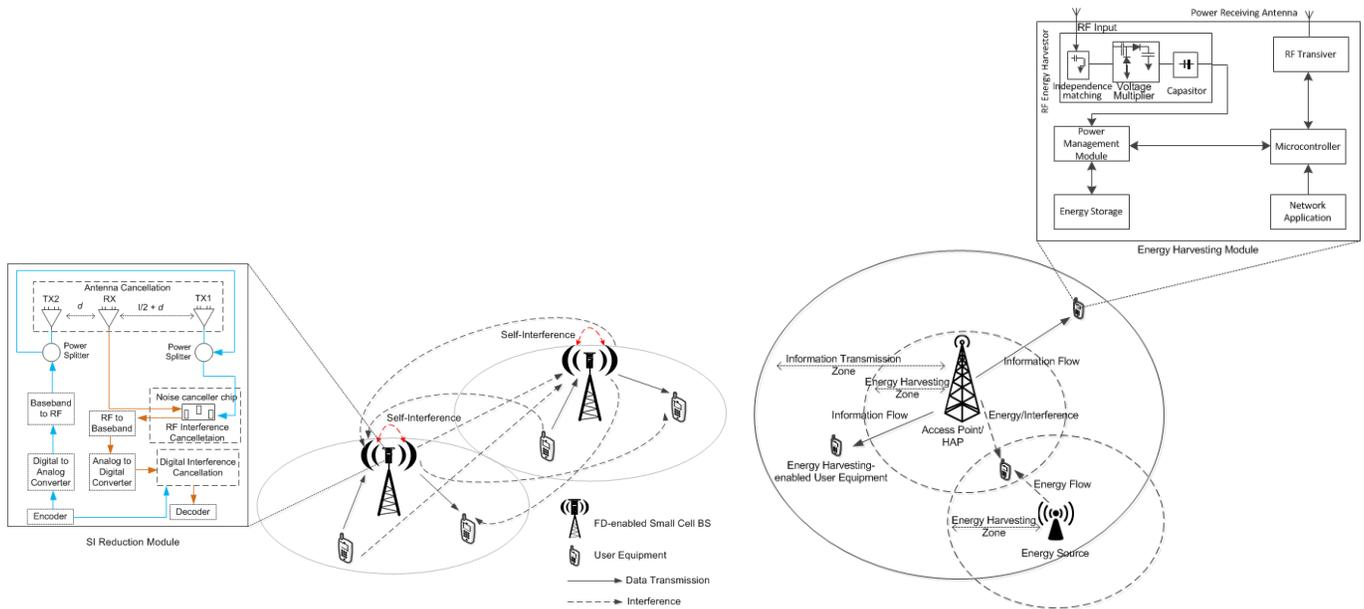

(b)  (c)

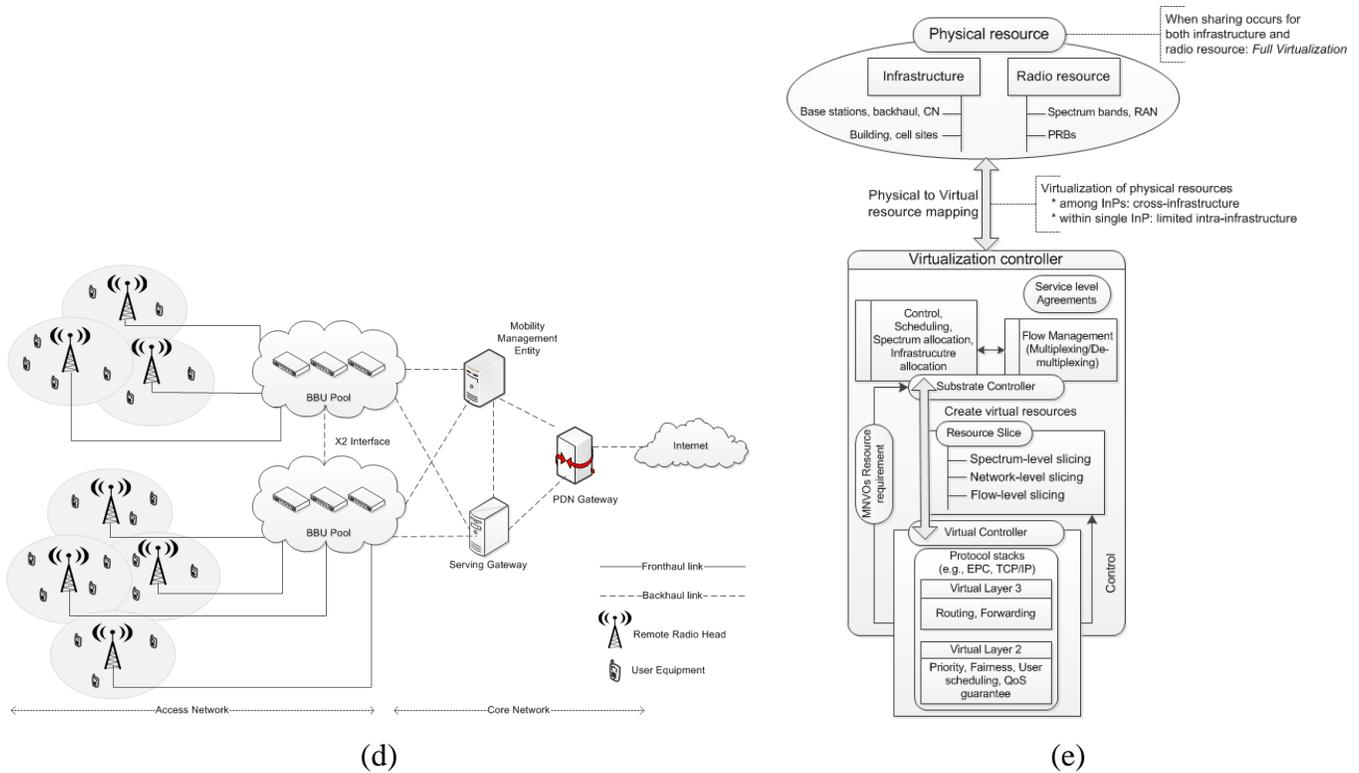

(d)    (e)

Fig. 1. (a) An illustration of conventional and 5G cellular wireless network. (b) FD-enabled small cell network. (c) Schematic diagram of an RF energy-harvesting network. (d) An LTE-A system enhanced with cloud-based radio access network. (e) A high-level representation of cellular network virtualization.

**Enabling Technologies for 5G Cellular Networks**

*Dense Heterogeneous Networks*

As shown in Fig. 1(a), the 5G cellular will be a *mulit-tier heterogeneous network* consisting of macrocells along with a large number of low power nodes such as (small cells, relays, remote radio heads [RRHs]) along with the provisioning for P2P (such as D2D and M2M) communication. The deployments of heterogeneous nodes in 5G systems will significantly have much higher density than today's conventional single-tier (e.g., macrocell) networks [5]. The heterogeneity of different classes of BSs (e.g., macrocells and small cells) provides flexible coverage area and improves spectral efficiency. By reducing the size of the cell, the area spectral efficiency is increased through higher spectrum reuse. Additionally, the coverage can be improved by deploying small cells indoors (such as home, office buildings, public vehicles etc.). Wireless P2P communication (e.g., D2D/M2M communication among UEs and autonomous sensors/actuators) underlaying cellular architecture can significantly increase the overall spectrum and energy efficiency of the network. In addition, the network-controlled P2P communications in 5G systems will allow other nodes (such as relay or M2M gateway), rather than the macrocell BS, to control the communications among P2P nodes. Given that the inter-tier and intra-tier interferences are well managed, the adoption of multiple tiers in the cellular network architecture will provide better performance in terms of coverage, capacity, spectral efficiency, and power consumption [6].

*Full-Duplex Communication*

In a FD communication scheme, a FD transceiver is capable of transmitting and receiving on the same frequency at the same time. It has been generally assumed that the wireless node (e.g., BS, UE etc.) cannot decode a received signal while it is simultaneously transmitting on a same frequency band due to internal interference between the transmitter and the receiver circuits (which is referred to as *self-interference* [SI]). However, with the recent advancements in the antenna and digital baseband technologies as well as the RF interference cancellation techniques, it is possible to build in-band FD radios. FD communication has the potential to double the spectral efficiency at the physical layer through the removal of a separate frequency band/time slot for both uplink and downlink transmission. Recent studies [7], [8] indicate that FD systems are feasible and can provide significantly higher data rates than the conventional half-duplex (HD) communication systems. FD technology can also solve the problems in existing wireless networks, such as hidden terminals, loss of throughput due to congestion, and large end-to-end delays [3]. For instance, FD communication scheme can reduce the latency by simultaneously receiving the feedback signals (i.e., channel state information [CSI], ARQ/ACK control signaling etc.) from the receiver during transmission. FD communication also enables a wireless node such as a BS to perform RF energy transfer [9] (e.g., wireless charging) while receiving uplink transmissions from UEs.

One possible approach to minimize SI in FD radios is to combine the *antenna cancellation*, *RF interference cancellation*, and *digital interference cancellation* techniques [7]. For instance, let the transmission signal be split between two transmit antennas. For a particular wavelength λ, two transmit antennas are placed at *d* and $\frac{\lambda}{2}+d$ away from the receive antenna. Hence by offsetting the two transmitters by half a wavelength, we can let the signals to add destructively. As a result, the receiver antenna receives a much weaker signal (e.g., less self-interference) compared with any one of the local transmit signal. After performing the antenna cancellation, the RF interference cancellation and the digital interference cancellation techniques can be employed to further decrease the self-interference. In RF interference cancellation technique, a *noise canceler chip* can remove a known analog interference signal from a received signal. Since the transmitted symbols are already known, a *coherent detection* mechanism is used in digital interference cancellation phase to reconstruct the signal.

Considering the small transmit power requirements to reduce the effect of SI, low-power networks (e.g., small cell networks [SCNs]) and short-range communications (e.g., D2D and M2M communications, multi-hop relaying) are potential scenarios where FD technology can be practically beneficial. However, in FD systems, interference management becomes significantly more complex due to new interference situations. For a multi-user channel-sharing situation, in addition to intra-cell interference (among the users in a cell), there are inter-cell downlink-to-uplink interference and inter-cell inter-user uplink-to-downlink interference.

*Energy-Aware Communication and Energy Harvesting*

One of the main challenges in 5G networks is to improve the energy efficiency of the battery-constrained wireless devices. In the context of prolonging the battery life and also improving the overall energy efficiency of the network, harvesting energy from energy sources could be an attractive solution. For instance, the UEs can harvest energy from environmental energy sources (e.g., solar and wind energy). However, due to stochastic nature of environmental sources, the available energy levels may vary significantly over time, locations, weather conditions etc. Therefore, harvesting energy from these sources may not be feasible for reliable and quality-of-service (QoS)-constrained wireless applications. Alternatively, energy can also be harvested from ambient radio signals (e.g., RF energy harvesting). In an RF-powered energy-harvesting network (RF-EHN), the UEs can harvest energy from hybrid access point (HAP) using RF signals for their information processing and transmission [9]. RF energy transfer is characterized by low-power and long-distance transfer, and thus suitable for cellular wireless environments (e.g., powering a large number of devices with relatively low energy consumption spreading in a wide area). Besides, the sustainable nature of RF energy sources make the RF-EHNs as a promising approach for future power/energy-constrained 5G wireless networks.

Generally an RF-EHN, as presented in Fig. 1(c), consists of RF energy sources, which can be either dedicated RF energy transmitters or ambient RF sources (e.g., TV towers). The access points (APs) refer to traditional network nodes (such as BSs, relays, SBSs etc.). Typically, the APs and RF energy sources have continuous and fixed electric supply. The harvesting-enabled UEs can harvest energy from RF sources. Depending on the network, the AP and RF energy source can be the same (e.g., acts as HAP). The AP has an energy harvesting zone and an information transmission zone. Since the operating power of the energy-harvesting component is much higher than that of the information decoding component, the energy harvesting zone is smaller than the information transmission zone [9].

As shown in Fig. 1(c), an RF energy-harvesting node (which could be user equipments [UEs], low-power femto APs etc.) consists of the following major components [9]: a low-power microcontroller, to process data from the network application; a low-power RF transceiver, for information transmission/reception; an energy harvester (composed of an RF antenna module, an impedance matching, a voltage multiplier, and a capacitor), to collect and convert RF signals into electricity. Besides, a power management module is used to decide whether the harvested energy should be used immediately or to store for future. The energy storage (e.g., battery) is used to reserve the harvested RF energy for future operation.

Since the amount of harvested RF energy depends on the distance from the RF source, the network nodes in the different locations can have significant difference in the amount of harvested RF energy [9]. For instance, when the energy harvester is distance $d$ apart from the transmitter with the unit transmit power, the amount of energy that can be harvested from the received signal is given by $E_h(d) = \varepsilon f(d, \alpha)$, where $f(d, \alpha)$ is the received power and $\varepsilon$ is the conversion efficiency, which depends on the efficiency of the antenna, the accuracy of the impedance matching between the antenna and the voltage multiplier, and the power efficiency of the voltage multiplier that converts the received RF signals to DC voltage. The received power function $f(d, \alpha)$ is given by $f(d, \alpha) = P_r^{det}(d) \times 10^L \times |r|^2$ where $P_r^{det}(d)$ is the received power using any deterministic free space path-loss model (e.g., Friis equation), path-loss factor $L = -\alpha \log_{10}(d/d_0)$ for a path-loss exponent $\alpha$ and reference distance $d_0$, and $r$ is a random number following complex Gaussian distribution.

*Simultaneous Information/Energy Transmission and FD-Enabled Energy Harvesting*

Since RF signals can carry both energy and information, theoretically RF energy harvesting and information reception can be performed from the same RF input signal. This scheme is referred to as the simultaneous wireless information and power transfer (SWIPT) [9]. Since information and power transfer operate with different power sensitivities at the receiver (e.g. –10 dBm and –60 dBm for energy and information receivers, respectively) [6], and the conventional receiver architecture is designed for information transfer only, practical SWIPT optimal circuits for harvesting energy are not available yet.

To implement SWIPT, one idea is to switch the receiver between two modes (e.g., information decoding and energy harvesting) based on instantaneous channel and interference condition. In the recent literature [9], the following receiver designs are available: *time switching* (e.g., an antenna periodically switching between information decoding and energy harvesting circuits) and *power splitting* (e.g. the receiver separates the received signal into two streams for the information and energy receivers with the same or varying power levels).

Leveraging the FD communication approach described before, the node can perform energy harvesting and data communication simultaneously. In cellular RF-EHNs, FD operations are typically supported with the use of dual-antenna HAPs [10]. However, this approach is different from conventional FD communication systems since one of those antennas performs downlink energy transfer to users and the other is used for information transmission in the uplink. Similar to any FD systems, self-interference (SI) is a critical issue for full-duplex HAPs, and there are ongoing research efforts (e.g., [10], [11]) to minimize SI in the HAPs.

*Cloud-Based Radio Access Network*

The concept of cloud-based radio access network (C-RAN) evolved from a distributed base station (DBS) architecture where a BS server is responsible for baseband processing. The baseband processing units (BBUs) of conventional cell sites are separated from the analog radio access units (referred to as *remote radio heads* [RRHs]) and moved to the *cloud* (e.g., BBU pool) for centralized signal processing and management. A BBU pool serves a particular area with a number of RRH of macro and small cells. Based on baseband signals received from the cloud, the transmissions of radio signals to users are performed by the RRHs.

In the 4G (e.g., LTE-A based cellular) architecture, radio and baseband processing functionality is integrated inside the BS and the inter-BS coordination is performed over X2 interface. In C-RAN, as shown in Fig 1(d), the baseband processing (such as coding, modulation, fast Fourier transform [FFT], etc.) is performed in the cloud (e.g., BBU pool). A BBU pool is a virtualized cluster, which can consist of general purpose processors to perform baseband processing. At a remote site, the radio unit (e.g., RRH) is co-located with the antennas and performs digital processing, digital to analog conversion, analog to digital conversion, power amplification and filtering [12]. The backhaul connects the cloud with the mobile core network. The fronthaul (generally optical transport link) spans from the RRHs to the cloud for transmitting digital baseband signals.

The advanced joint processing mechanism in the cloud can reduce the delay during intra-BBU pool handover in conventional BSs. The use of inexpensive and densely deployed RRHs in C-RAN will enhance the scalability, improve network capacity, and extend the coverage of future 5G systems. C-RAN architecture also significantly reduces the capital expenditure (CAPEX) and operational expenditure (OPEX). For example, C-RAN lowers the cost of baseband processing and reduces the power consumption by performing load balancing and cooperative processing of signals from several BSs. The number of BS sites can be greatly reduced by means of BBU aggregation in the cloud, which results in much lower cost of operation. In addition, C-RAN minimizes energy consumption by enabling the MIMO or coordinated multipoint (CoMP) concepts. The native centralized support for multi-standard operation in C-RAN will make inter-RAT operation (e.g., scheduling, interference coordination or traffic management) and network maintenance (e.g., upgrade to newer technology) relatively easy [12].

*Wireless Network Virtualization*

Cooperation and network virtualization will become one of the main trends in 5G systems. Together with C-RAN architecture, wireless network virtualization (WNV) facilitates resource sharing among many operators. WNV enables multiple network operators to share common resources (e.g., network infrastructure, backhaul, licensed spectrum, core and radio access network [RAN], energy/power etc.). The virtualization mechanism abstracts (e.g., isolates) the physical resources to a number of virtual resources, which is shared by different consumers (e.g., service providers). The main advantages of WNV include: high resource utilization, improved system performance, reduced CAPEX and OPEX, better quality-of-experience (QoE) for the end users, and easier migration to newer technologies by isolating part of the network [13].

The virtualized wireless networks consist of an *infrastructure provider* (InP) and *mobile virtual network operator* (MVNO). The InP owns the physical cellular infrastructure and radio resources. The MVNO leases the resources from InP, creates and operates virtual resources, and assigns to the subscribers. The network resources (e.g., infrastructure, radio spectrum) belong to one or more InPs and are virtualized and spitted into *slices*. The MVNO utilizes the slices from InP depending on specific service level agreements (SLAs) and provides service to the end users without knowing the undelaying physical network architecture.

*Physical and virtual resource sharing in WNV*: Depending on different degrees of virtualization, the WNV can be classified as: (i) *Cross-infrastructure virtualization:* enables to virtualized resources across InPs (inter-InP-virtualization) and within InPs (intra-InP-virtualization). Hence all of the InPs in a geographical region can share the network resources across the MVNOs. (ii) *Limited intra-infrastructure virtualization:* considers resource virtualization within a single InP, e.g., radio spectrum and access network is shared by different MVNOs.

A high level representation of WNV is illustrated in Fig. 1(e). The physical network sharing refers to the sharing of radio spectrum and/or wireless network infrastructure. Radio spectrum resources refer to licensed or dedicated free spectrum bands, subchannels, or physical resource blocks (PRBs) for LTE-A networks. Using spectrum sharing, all or part of the spectrum owned by the InP are utilized by one or multiple MVNOs. Infrastructure sharing can be categorized as *active* and *passive* sharing [13]. Active sharing refers to sharing of network elements (e.g., RF antennas, BSs, backhaul and backbone, routers and switches in the core network etc.). The passive sharing is the sharing of non-networking materials (e.g., buildings, locations, cell sites etc.). The infrastructure sharing offers cross-infrastructure virtualization, e.g., multiple InPs can share the same physical network.

Virtual wireless resources are created by slicing the physical network resources (e.g., spectrum and/or infrastructure) into virtual pieces. For example, an MVNO may have the core network but no radio coverage for a specific region, and may request RAN slice from the InP. The resource manager or virtualization controller (often referred to as *hypervisor*) is responsible for mapping the physical to virtual resources. There are two parts of the controller, *substrate controller* and *virtual controller* [13]. The substrate controller allows InPs to virtualize and manage the substrate physical network where the virtual controller is used for MVNOs to manage the virtual slices or networks. Using the virtual controller, the MVNOs can customize their own networks and services.

The virtual resources can be sliced into different levels as follows [13]: (i) *Spectrum-level slicing:* the radio resources are sliced and assigned to MVNOs by time/space multiplexing, or spectrum reuse method (e.g., underlay accessing). The hypervisor schedules the air interface resource (e.g., PRBs for LTE-A/LTE-B based network) between BS and UEs. (ii) *Network-level slicing:* it allows to virtualize the physical network nodes (e.g., BSs, relays, femto APs), core networks (e.g., mobility management entity [MME], serving gateway [SGW]) and computing modules (e.g., CPU, memory, I/O devices) in a close geographical area into virtualized nodes based on some criterion (e.g., users link quality, MNVOs' resource requirements, power budget, interference etc.). (iii) *Flow-level slicing:* In flow-level slicing, the resource slice is defined as the set of virtual resources (e.g., traffic flows) requested by the MVNOs. The resource slice could be based on bandwidth (e.g., data rate) and/or radio resource (e.g., time slot) and the cellular network topology is transparent to the MVNOs.

**Research Challenges for 5G Networks**

Despite the fact that aforesaid key-enabling technologies significantly improve the overall network performance and end-user QoE, there are several research challenges (from system/device design and testing to network management) to fulfill the requirements of 5G systems. In the following, we discuss the research challenges into two categories, namely, (i) measurement and test challenges for future 5G systems; and (ii) challenges for efficient management of the radio resources.

*Measurement and Test Challenges for 5G Systems*

In the following, we briefly present the key challenges related to measurement, testing, and validating the performance of 5G system components. Table II summarizes the issues that make the measurement and testing of 5G systems more challenging.

*Measurement and modeling of the propagation channels*: The fundamental challenge to measure and model the propagation channel is mainly due to use of higher frequencies and various bandwidths, together with much larger antenna arrays. The growth of wireless traffic requires additional spectrums (e.g., higher frequencies in the mm-wave) and the enabling technologies such as large-scale MIMO and densification of network nodes impose new requirements for channel measurement/modelling in the spatial domain. The key challenges/considerations for measuring and modelling the 5G propagation channels can be summarized as follows [14].

- *Efficient and realistic measurement*: Since the measurement data are crucial for the needed extensions/modifications of available propagation models, the measurement approach should capture diverse frequency range, spatial consistency, 3D (e.g., elevation) and spherical waves, as well as the new communications paradigms such as D2D/M2M and small cell communications. Besides, measurements should be carried out for mm-wave (e.g., 60 GHz and above) both in indoor/outdoor and considering possible realistic use-cases (such as crowded areas, vehicle-to-vehicle/roadside communications etc.).
- *Spatial distributions and mobility*: The existing channel models are *drop-based*, e.g., the scattering environment is randomly created for each link. However, since the density of links is expected to increase in 5G systems, it is important to model these links in a consistent manner which can also inherently support the heterogeneous mobility behavior of different network nodes.
- *Large-scale antenna arrays and mm-wave communication*: It is envisioned that 5G systems will use very large-scale antenna arrays for directive communication. Considering these directive antennas, current channel models require corresponding improvement in angular resolution as well as sub-path amplitude distribution. In addition, these large-scale antenna arrays need *spherical wave* modelling instead of the commonly used *plane wave* approximation [14]. The use of mm-wave frequencies (particularly at 60 GHz) is proposed to achieve spectrum and spatial multiplexing gain. However, the characteristics such as highly resolved angular properties and non-line-of-sight (NLOS) path-loss are not well known; and hence require further measurement studies.

*Testing 5G systems:* The use of wide channel bandwidths, high data rate requirements, fast response times, more complex antenna configurations, and support for multiple radio access technology (RAT) pose significant challenges to the development of next generation BSs and devices. Testing of 5G systems is not limited to the hardware only, but it will be also required to validate new communications algorithms and approaches.

From the perspective of an RF engineer, some of the key design and measurement challenges for 5G systems are as follows [15, Ch. 6]: (i) the requirement to handle multiple channel bandwidths (e.g., 1.4 to 20 MHz for LTE-A), operability in higher bands (e.g., 60 GHz for mm-wave); (ii)

The use of different transmission modes in LTE-A (e.g., orthogonal frequency-division multiple access [OFDMA] for uplink, single-carrier frequency-division multiple access [SC-FDMA] for downlink) and also provisioning for both transmission modes (e.g., time/frequency-division duplexing [TDD/FDD]); (iii) support for multi-RAT (e.g., due to multi-standard radio transmission requirements) and simultaneous transmission/reception of control, energy and data; (iv) the spectral, power, and time variations (due to heterogeneous network traffic and node density), and need to support multi-antenna techniques (e.g., transmitter diversity, spatial multiplexing, beamforming). The issues related to measurement and testing of the receiver radios for the 5G systems may include [15, Ch. 6]: (i) verifying the RF and baseband receiver (e.g., autonomic gain control, baseband demodulation and hybrid automatic repeat request [HARQ] functional testing); (ii) receiver performance under impaired conditions (e.g., phase noise and additive white Gaussian noise [AWGN] impairments); (iii) receiver performance testing (e.g., testing under channel propagation impairments); (iv) testing the MIMO-enabled systems (e.g., baseband coding assessment, receiver verification challenges such as isolation of signals, performance testing under static and faded channels). There are also challenges related to testing the 5G BSs and UEs as discussed below.

- *Challenges of testing the 5G BSs*: 5G systems will be able to re-use some of the legacy infrastructure in the existing core network. However, the 5G air-interface specifications will include a range of new features and concepts, and thus require a significant development program. These features may include massive MIMO, low-latency HARQ procedures, high-order modulation schemes (e.g., 256 quadrature amplitude modulation [QAM]), the broad set of combinations/configurations of RF bands and channel bandwidths to provide the spectrum flexibility, testing FD radios, and energy harvesting capabilities. As opposed to conventional systems, future BSs (including small cell BSs, relays etc.) will also need to handle radio resource control (including admission control, load balancing and radio mobility decisions) on top of the traditional Layer 1 and Layer 2 functionalities. In addition, the *stress testing* should be conducted for the BSs, clouds, and the core networks to validate the performance against large numbers of network nodes.

- *Challenges of testing the 5G UEs*: The testing issues for 5G UEs are similar to those for traditional systems using measurement matrices such as maximum output power, power control, and receiver sensitivity. However, the need for the use of OFDMA transmission scheme in the downlink and SC-FDMA in the uplink in LTE-A/LTE-B based 5G systems, as well as support for simultaneous connections with provisioning for energy harvesting capabilities require new measurement concepts to support the necessary tests. For proper RF measurements, the test instrument should incorporate a signaling protocol that operates automatically with user definable parameters (such as channel number). The *functional testing* for the UEs should incorporate testing signaling protocol, end-to-end throughput, and handover testing. One of the principal challenges for testing 5G UEs is to ensure that the state change response requirements are met. For example, in *idle* mode, the device will be in a low power consumption state to ensure good battery life, and only periodically check for paging messages. When the device is in *connected* mode (e.g., data transmission is to be scheduled), the device must wake up and rapidly synchronize its uplink/downlink (or even both in case of FD communication). To ensure energy efficiency, *battery drain testing* should be performed in all the phases (e.g., product development, design validation, and application design) of system development cycles [15, Ch. 6].

TABLE II

MEASUREMENT AND TEST ISSUES FOR 5G SYSTEMS

| Particulars | Issues to consider |
| --- | --- |
| Channel measurement and modeling | Massive number of devices, heterogeneous traffic pattern and node density, support for spatial diversity (e.g., massive-MIMO), operability in higher bands, new use cases (e.g., D2D/M2M communication) |
| Testing RF modules, transmitters and receivers | Testing support for MIMO and beamforming capabilities, testing simultaneous transmission/reception and FD radios, testing under different propagation bands and under impairments, cost effectiveness (testing large number of units with low cost) |
| Testing BSs and UEs | Stress testing with large number of nodes, various traffic |

| | patterns etc., performance testing with standard (e.g., LTE-A) specifications, flexibility to operate with multi-standard, energy efficiency testing for UEs and low power nodes (e.g., small cell BSs, relays, RRHs etc.) |
|---|---|

*Research Challenges from the Perspective of Radio Resource Management*

The following section briefly outlines some of the fundamental research challenges in the context of radio resource management (RRM) in future 5G networks. Table III summarizes the open research issues.

*Interference management in heterogeneous networks*: Due to the dense deployment of heterogeneous nodes in 5G networks, one approach of improving the resource utilization to use the available resources as a spectrum underlay manner. However, for the underlay communication 5G networks, interference management is one of the key challenges. One of the biggest challenges for muti-tier heterogeneous networks is to mitigate inter-cell and inter/intra-tier interferences. In addition to heterogeneity and dense deployment of wireless devices, coverage and traffic load imbalance due to varying transmit powers of different BSs make the interference management and resource allocation problems more challenging than those in conventional single-tier systems. Besides, different access restrictions (e.g., public, private, hybrid etc.) in different tiers lead to diverse interference levels. The priorities in accessing channels necessitate the adaptation of different dynamic resource allocation strategies. The introduction of carrier aggregation, cooperation among BSs (e.g. CoMP), and the provision of P2P communication also complicate the dynamics of the interference. Besides, in order to improve the overall spectrum efficiency, the network nodes need to select the offloading (e.g., UEs in macro-tier may offload to SCNs) and mode selection (e.g., direct, D2D, relay-aided D2D etc.) decisions dynamically based on network condition (e.g., load, channel states, interference etc.). Since a large number of devices in 5G systems will reuse the same frequency, setting up (e.g., device discovery) and maintaining links for P2P communications could be difficult. Due to the huge number of devices and their massive accesses (especially for M2M communications), existing medium access control (MAC) mechanisms will need to be redesigned.

*Full-duplex communication*: In the context of FD communication, SI limits the theoretical gain of the FD transmission. Even with the several SI cancellation mechanisms, a certain amount of SI remains in the system. To achieve the performance limit of FD transmission, efficient SI cancellation techniques are required that can eliminate residual SI. As has been mentioned before, there are new sources of inter-cell and intra-cell interference in FD communication due to simultaneous transmission/receptions. Moreover, due to high transmit power of BSs, downlink-to-uplink interference becomes more severe compared to the uplink-to-downlink interference. Hence, to achieve maximal spectral efficiency, advanced channel estimation, interference management, and power allocation mechanisms (utilizing advanced techniques such as transceiver beam-forming, distributed space-time coding etc.) will be required considering new interference sources. Besides, the FD transceivers should able to opportunistically select the transmission modes (e.g., HD/FD) depending on the network dynamics.

*Energy-harvesting communication*: For RF-EHNs, efficient wireless charging as well as channel, power, and time adaptation policies will need to be investigated. Conventional multiuser scheduling schemes (e.g., greedy/opportunistic/round-robin etc.) do not consider the amount of harvested energy and/or energy requirements, and may lead to energy (or transmit power) outages and/or energy overflows in RF-EHNs. Therefore, efficient multi-user scheduling schemes will be required for RF-EHNs. Also, efficient CSI acquisition techniques are required at HAPs to optimize the transmit power or energy harvesting duration to perform directional energy transfer depending on the channel conditions. Besides, the issues such as distributed energy beamforming, interference management, and effect of the mobility of devices/energy sources need to be investigated.

*Cloud-RAN*: There are various challenges to deploy C-RANs, such as optimally utilizing the processing resource, efficiently using the fronthaul links connecting BBUs with RRHs, and centralized control of the propagation signal. The performance of C-RAN is constrained by the limited fronthaul capacity. Hence fronthaul-aware resource allocation schemes (including efficient signal quantization/compression and beamforming techniques) are required to maximize the network throughput. To achieve optimal energy savings of the C-RAN, RRHs need to be served optimally considering the number of active BBUs in the BBU pool and network load (e.g., traffic demand from UEs/RRHs). The existing resource sharing algorithms need to be enhanced considering clustering the cells (e.g., RRHs) and BBUs (in the cloud) to reduce scheduling complexity.

*Wireless network virtualization*: A fully-functional WNV framework requires efficient resource utilization, inter-slice isolation, and customizable intra-slice resource allocation. The resource allocation algorithms need to be designed considering the finite and heterogeneous (e.g., spectrum, infrastructure, computing) resources, resource requirements of MVNOs, fairness criteria etc. Besides, control signaling and bootstamping issues need to be addressed for spectrum virtualization. The issues related to resource discovery, isolation, pricing-based allocation, and mobility management (e.g., seamless handover to different MVNO/InP) open up new research opportunities.

TABLE III

SUMMARY OF EMERGING TECHNOLOGIES FOR 5G CELLULAR NETWORKS

| Enabling technologies and trends | Benefits and features | Applicability | Fundamental research challenges |
|---|---|---|---|
| Heterogeneous multi-tier networks | Increased throughput, spectrum utilization, energy efficiency, coverage expansion | Small cell networks, D2D/M2M communication, Internet-of-Things | Interference management, adaptive power control, dynamic mode selection and offloading to underlay network, device discovery, unified MAC design |
| Full-duplex communication | Spectrum efficiency, reduced latency, energy efficiency | Small cell networks, D2D communication, cognitive radio networks, multi-hop relaying | SI reduction, cross-layer resource management, power allocation, interference management, synchronization and time adjustment to establish FD transmission, dynamic mode selection, designing a MAC protocol |
| Energy harvesting networks | Energy efficiency, wire free and energy-aware (green) | Small cell networks, | Muti-user scheduling, advanced channel acquisition, energy |

| | communication | D2D/M2M communication | beamforming, harvest/transmit time adaptation, interference management, SWIPT-enabled resource allocation |
| --- | --- | --- | --- |
| Cloud RAN | Scalability, energy/power savings, increased throughput, reduced delay, adaptability to dynamic traffic, reduced CAPEX/OPEX, easier network management | Service oriented communication, heterogeneous networks | BBU management (e.g., cooperation, interconnection, clustering), energy-aware scheduling, fronthaul-aware resource allocation |
| Wireless network virtualization | Resource utilization, improved throughput, energy savings, reduced CAPEX/OPEX, enhanced QoE, easier migration/maintenance | Service oriented communication (everything as a service) | Isolation, resource allocation, fairness, revenue/price optimization, mobility management |

**Conclusion**

We have provided an overview of several emerging technologies for 5G cellular wireless networks. Some of the open research problems have been outlined including those related to testing and measurement of 5G systems. In addition to the technologies discussed above, technologies such as mm-wave [3] and massive-MIMO [3], [4] will also impact the design and development of 5G networks. Future 5G cellular wireless networks will definitely be a combination of different enabling technologies. However, the biggest challenge will be to integrate all the enabling technologies and make them all work together.

**Ekram Hossain** (F'15) is a Professor (since March 2010) in the Department of Electrical and Computer Engineering at University of Manitoba, Winnipeg, Canada. He received his Ph.D. in Electrical Engineering from University of Victoria, Canada, in 2001. Dr. Hossain's current research interests include design, analysis, and optimization of wireless/mobile communications networks, cognitive radio systems, and network economics. He has authored/edited several books in these areas (http://home.cc.umanitoba.ca/~hossaina). Dr. Hossain serves as the Editor-in-Chief for the *IEEE Communications Surveys and Tutorials* and an Editor for *IEEE Wireless Communications*. Also, he serves on the IEEE Press Editorial Board. Previously, he served as the Area Editor for the *IEEE Transactions on Wireless Communications* in the area of ``Resource Management and Multiple Access" from 2009-2011, an Editor for the *IEEE Transactions on Mobile Computing* from 2007-2012, and an Editor for the *IEEE Journal on Selected Areas in Communications – Cognitive Radio Series* from 2011-2014. Dr. Hossain has won several research awards including the University of Manitoba Merit Award in 2010 and 2014 (for Research and Scholarly Activities), the 2011 IEEE Communications Society Fred Ellersick Prize Paper Award, and the IEEE Wireless Communications and Networking Conference 2012 (WCNC'12) Best Paper Award. Dr. Hossain was elevated to an IEEE Fellow "for contributions to spectrum management and resource allocation in cognitive and cellular radio networks". He is a Distinguished Lecturer of the IEEE Communications Society (2012-2015). He is a registered Professional Engineer in the province of Manitoba, Canada.

**Monowar Hasan** (S'13) is currently working toward his M.Sc. degree in the Department of Electrical and Computer Engineering at the University of Manitoba, Winnipeg, Canada. He has been awarded the University of Manitoba Graduate Fellowship. Monowar received his B.Sc.

degree in Computer Science and Engineering from Bangladesh University of Engineering and Technology (BUET), Dhaka, in 2012. His current research interests include Internet of things, wireless network virtualization, and resource allocation in 5G cellular networks. He served as a reviewer for several major IEEE journals and conferences.